\documentclass[letterpaper,10pt]{article} 
\usepackage{opticameet3} %% use version 3 for proper copyright statement
\usepackage{booktabs}
\usepackage{colortbl} % for shading table

\newcommand\authormark[1]{\textsuperscript{#1}}

\usepackage{amsmath,amssymb}
\usepackage[colorlinks=true,bookmarks=false,citecolor=blue,urlcolor=blue]{hyperref} %pdflatex
\usepackage{physics}

\begin{document}
\title{Collection of fluorescence from an ion\\using trap-integrated photonics}

% \author{Author name(s)}
% \address{Author affiliation and full address}
\author{Felix W. Knollmann,\authormark{1,*} Sabrina M. Corsetti,\authormark{1} Ethan R. Clements,\authormark{1} Reuel Swint,\authormark{2} Aaron D. Leu,\authormark{1,3} May E. Kim,\authormark{2}  Patrick T. Callahan,\authormark{2} Dave Kharas,\authormark{2} Thomas Mahony,\authormark{2} Cheryl Sorace-Agaskar,\authormark{2} Robert McConnell,\authormark{2} Colin D. Bruzewicz,\authormark{2} \\ Isaac L. Chuang,\authormark{1} Jelena Notaros,\authormark{1} and John Chiaverini\authormark{1,2}}
\address{\authormark{1} Massachusetts Institute of Technology, Cambridge, MA 02139, USA\\
\authormark{2} Lincoln Laboratory, Massachusetts Institute of Technology, Lexington, MA 02421, USA\\
\authormark{3} University of Oxford, Oxford OX1 3PU, United Kingdom}
\email{\authormark{*}fwk@mit.edu} 

\begin{abstract*}
Spontaneously emitted photons are entangled with the electronic and nuclear degrees of freedom of the emitting atom, so interference and measurement of these photons can entangle separate matter-based quantum systems as a resource for quantum information processing. However, the isotropic nature of spontaneous emission hinders the single-mode photonic operations required to generate entanglement. Current demonstrations rely on bulk photon-collection and manipulation optics that suffer from environment-induced phase instability, mode matching challenges, and system-to-system variability\textemdash factors that impede scaling to the large numbers of entangled pairs needed for quantum information processing.  To address these limitations, we demonstrate a collection method that enables passive phase stability, straightforward photonic manipulation, and intrinsic reproducibility. Specifically, we engineer a waveguide-integrated grating to couple photons emitted from a trapped ion into a single optical mode within a microfabricated ion-trap chip. Using the integrated collection optic, we characterize the collection efficiency, image the ion, and detect the ion’s quantum state. This proof-of-principle demonstration lays the foundation for leveraging the inherent stability and reproducibility of integrated photonics to efficiently create, manipulate, and measure multipartite quantum states in arrays of quantum emitters.
\end{abstract*}

\section{Introduction}
Individually confined atoms or ions form an effective platform for storage, processing, and measurement of quantum states due to their controllability and isolation from the environment~\cite{colin}.  Additionally, their abundance of optical transitions provides a mechanism to convert quantum information between matter and photonic states. This mechanism enables high-fidelity state preparation and measurement, as well as coherent, non-local connectivity.  The detection of emitted single photons allows the projection of internal atomic states, but further manipulation of photon properties at the quantum level calls for single-mode guiding structures. The challenge of collecting isotropically emitted light into such single-mode structures limits the effectiveness of protocols generating entangled states for sensing, communication, or distributed quantum computing.  Transferring emitted photons into single-mode optical fiber using bespoke, free-space lenses~\cite{Monroe2007,oxford_reg,duke_fast_reg} or high-finesse cavities~\cite{simon_cavity,innsbruck_cavities} has enabled proof-of-principle demonstrations of remote entanglement generation between trapped ions or atoms.  However, the inherent challenges to robustness, phase and mode stability, and extensibility in bulk-optical approaches make them unlikely to allow high-fidelity creation of novel multipartite entangled states of many atomic systems~\cite{gabriel_multipartite} or rapid coupling of large arrays of qubits in separate quantum information processing modules~\cite{PME}.

Incorporating integrated optics may improve upon bulk optical approaches for collection of spontaneous emission. Since ions can be trapped tens of microns above the chip surface in a microfabricated trap~\cite{surface_trap2005}, a grating integrated in that surface can achieve a numerical aperture similar to that of a free-space optic while minimizing lateral footprint—making this approach inherently scalable and well-suited for extensible quantum architectures. Each grating could collect light into a single-mode waveguide for integrated interference and photon-mediated entanglement. Alternatively, gratings could direct light to a waveguide-coupled detector placed far from the ion to allow readout via resonance fluorescence~\cite{Wineland1998} while mitigating adverse effects arising from the interplay of trapping and detector fields~\cite{susanna_snspd,dave_spad,Setzer2021}.  Furthermore, tailoring the spatial profile of collected light using multiple phase-stable integrated collecting optics can potentially mitigate sources of entanglement infidelity related to the finite temperature of confined atoms and ions~\cite{cornell,david2025recoil}.  Initial demonstrations using integrated photonic waveguides to \textit{deliver} control light to trapped ions already highlight the benefits in path-length and beam-pointing stability when compared to traditional free-space optics~\cite{Mehta2016,Mehta2020,Niffenegger2020,sandia_array2024,pgc_ions}. 

In this work, we collect ion fluorescence into a trap-integrated single-mode waveguide to demonstrate the key requirement for integrated photon-mediated entanglement. Because of the reciprocity of electromagnetic-wave propagation through passive structures, designing for light collection from a small volume can be reframed as engineering a tightly focused emitted beam. To design a tightly focusing integrated optic, we introduce a novel method for tailoring the scattering strength of a diffraction grating to match a desired emission profile while also increasing the effective grating length and minimizing aperture-induced imperfections.  We co-fabricate the grating with an ion trap and image its emission profile. By collecting photons emitted from the ion into a single-mode waveguide, we experimentally verify a model of grating collection efficiency based on ion position and emitted-photon polarization.  As a proof of principle, we use the integrated collection path to detect 422-nm light from a \(^{88}\)Sr\(^+\) ion to determine it's quantum state.  This demonstration is an initial step toward high-performance remote entanglement generation and state detection of trapped ions with integrated collection optics. The demonstrated technique applies to all ion species of practical interest for quantum information processing and sensing and could immediately be extended to trapped neutral atoms and other point source emitters. The following presents the grating design, characterization, and measurement, and then discusses the implications of our results.

\begin{figure}[t]
  \centering
\includegraphics[width=0.9\textwidth]{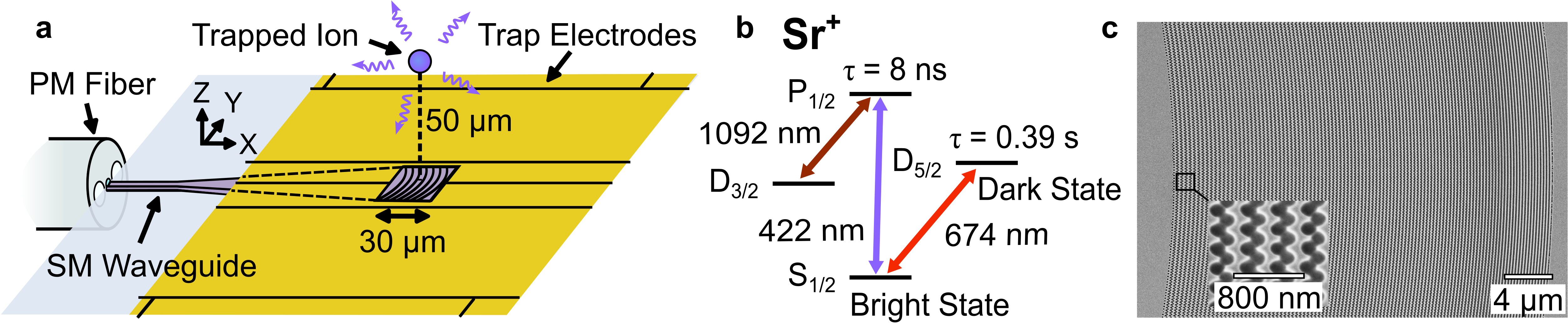}
\caption{A focusing diffraction grating collects fluorescence photons from a trapped ion into a single-mode (SM) waveguide. \textbf{a} The schematic shows the grating's placement beneath the trap electrodes and the path through a non-linear taper to convert from the 30-\textmu m grating width to a single-mode waveguide, which routes to an edge-coupled fiber. The ion height above the trap surface is 50 \textmu m (not to scale). \textbf{b} A simplified level structure shows the relevant transitions and excited state lifetimes. \textbf{c} The inset in this scanning-electron micrograph of the collection grating shows the detailed structure of phase-shift apodization.}
\label{fig:overview}
\end{figure}

%%%%%%%%%%%%%%%%%%%%%%%%%%%%

\section{Results}
For our proof-of-principle demonstration (see Fig.~\ref{fig:overview}), we fabricate ion traps with an integrated photon collection device using a 200-mm wafer-scale process that offers low losses at visible wavelengths~\cite{platform}. The photonics platform consists of two layers of 100-nm-thick silicon nitride used for the grating and a layer of 100-nm-thick alumina used for low-loss photon routing; 90 nm of SiO\(_2\) separate each layer.  The waveguide layers are clad above and below with 5-$\mu$m-thick SiO\(_2\) layers.  As shown in Fig.~\ref{fig:overview}a, we deposit and etch a layer of niobium above the upper cladding layer to define electrodes for RF and DC voltages that confine a single \(^{88}\text{Sr}^+\) ion 50 \textmu m above the chip surface (see simplified level structure in Fig.~\ref{fig:overview}b). A 50-nm-thick transparent conductive film (indium-tin oxide) covers the dielectric (SiO\(_2\)) exposed in gaps between metal electrodes above the grating.  A quantizing magnetic field of 4~G is applied in a direction normal to the trap surface.

%%%%%%%%%%%%%%%%%%%%%%%%%%%%%%%%%%%%%%%%%%%%%%%%%%%%%%%%%%%%%%%%%

\subsection{Grating device design}
\begin{figure}[t]
  \centering
\includegraphics[width=0.65\textwidth]{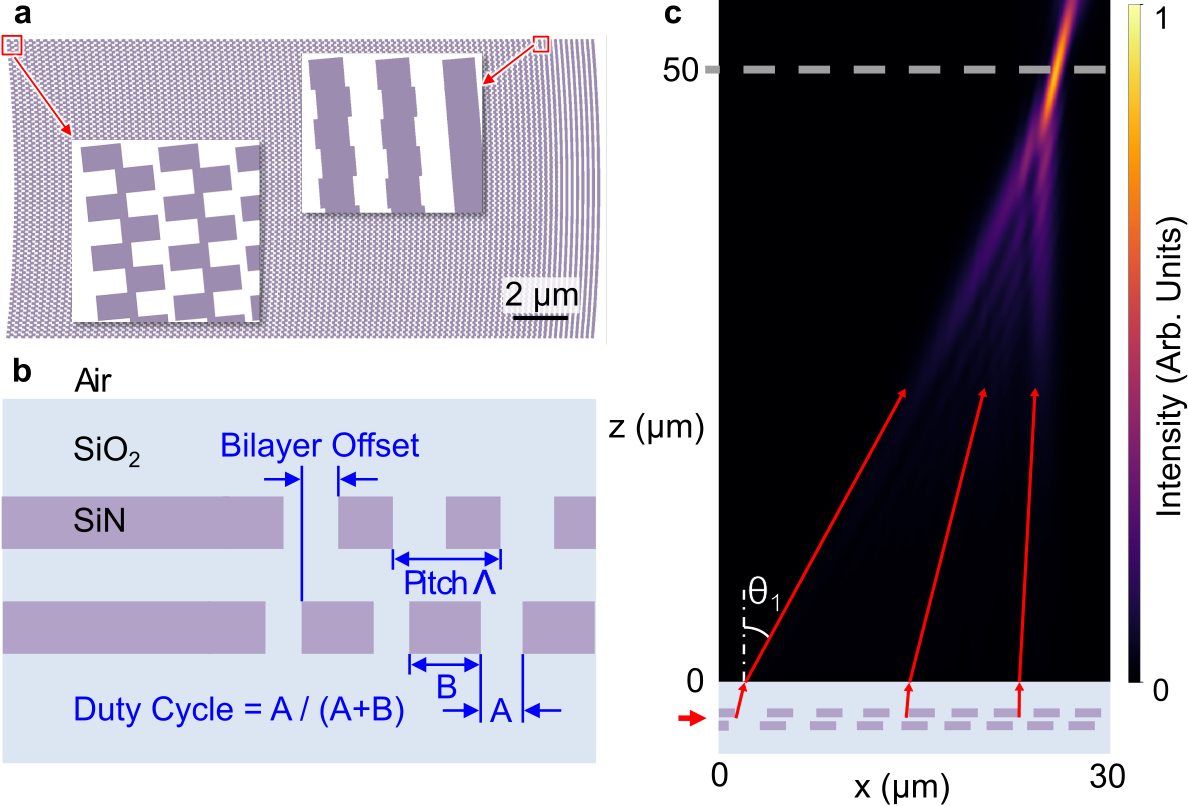}
\caption{Dual-layer apodized diffraction grating design. \textbf{a} A focusing grating designed using phase-shift apodization. The insets show details of the phase-shift apodized grating teeth near the leading and trailing edge of the grating (see Methods subsection~\ref{psa_descr} for details).  \textbf{b} Cross-sectional structure of the dual-layer grating with materials indicated. Varying the grating pitch, \(\Lambda\), provides longitudinal focusing (as shown in panel \textbf{c}).  The bilayer offset breaks the vertical symmetry to obtain high upward directivity.  \textbf{c} The vertical slice of the simulated grating emission shows longitudinal focusing to a small spot at the targeted ion height of 50 \textmu m.}
\label{fig:grating_design}
\end{figure}

Designing and fabricating photonic devices to efficiently collect photons from an atomic point source at a working distance of tens of microns is challenging because it requires both low losses and sub-wavelength feature sizes at the short wavelengths relevant for ion transitions~\cite{platform}. The collection efficiency can be mapped to the projection of the normalized electric field emitted by the optic onto the ion's radiating-dipole emission pattern~\cite{cornell}. Assuming constant diffraction efficiency, minimizing the focused spot size at the ion height thus maximizes the collection efficiency. Therefore, we optimize the grating parameters visualized in Fig.~\ref{fig:grating_design} to create a tight focus by varying the grating pitch and setting the curvature of the grating teeth to concentrate rays both longitudinally and transversely~\cite{Cheng2020,gillen_grating}. Furthermore, maximizing upward directivity maximizes the fraction of light emitted toward the ion and prevents uncontrolled reflections from the lower interfaces in the chip structure that would not optimally couple to the ion. We achieve upward directivity by designing our gratings using two waveguide layers with a relative lateral shift in the propagation dimension [Fig.~\ref{fig:grating_design}b] to break the vertical symmetry~\cite{karan_grating_design,gillen_grating,Corsetti2023,Notaros_bilayer2016}. Finally, we tailor the grating amplitude profile to maximize the overlap with the ion emission incident on the grating, as depicted in Fig.~\ref{fig:grating_design}a,c. We describe the full optimization sequence in the Methods section.

Existing techniques for apodizing (tuning the local grating strength) to tailor the grating amplitude profile over a large grating size include varying the duty cycle of the etch within a single period or varying the depth of the etch~\cite{apodized_grating_design,apodized_grating_alt}. However, gratings designed using these existing techniques can be difficult to fabricate, as they require feature sizes substantially below the optical wavelength and most relevant electronic transitions in ions have short wavelengths. We have therefore developed a novel apodization method that enables continuous variation of the grating strength up to a maximum value without the need for feature sizes inaccessible to typical photolithography. Our technique subdivides the grating into sub-wavelength zones just under half a period in width (Fig.~\ref{fig:grating_design}a). Shifting every other zone longitudinally by a continuously tunable distance shifts the phase of the locally emitted light such that scattering from neighboring zones partially or fully destructively interferes. We describe the detailed implementation in Methods subsection~\ref{psa_descr}. The only fundamental requirement is that the transverse period must be less than the wavelength of the light in the medium to prevent transverse diffraction.  Using this technique, which we call phase-shift apodization, we design a grating with an emission profile that strongly suppresses extraneous longitudinal beam structure and closely matches the ion radiation mode (see Fig.~\ref{fig:grating_design}c).

The applications of photon-mediated entanglement and state determination pose different constraints on the grating design based on the relevant atomic physics. While the signal-to-noise ratio of ion fluorescence to background scatter of the applied driving light is crucial to the rate and fidelity of state determination~\cite{colin,Oxford_MLE}, it is of minimal importance to photon-mediated entanglement where the ion emits a photon while all lasers are off. Improving the coupling of the grating to the ion not only improves the signal but may also reduce the noise by making it less likely that background scatter couples into the waveguide mode. The second important difference lies in the polarization behavior. While state determination often depends on fluorescence from a combination of atomic transitions with different polarizations, photon-mediated entanglement always relies on coupling specific photon states with specific atomic states and is thus transition and polarization specific~\cite{cornell,Monroe2007}.  Here we aimed to design a grating optic with as tight a focus as possible (for one polarization) while observing a constraint on the minimum achievable feature size.
%%%%%%%%%%%%%%%%%%%%%%%%%%%%%%%%%%%%%%%%%%%%%%%%%%%%%%%%%%%%%%%%%

\subsection{Grating device characterization}
\begin{figure}[t]
  \centering
\includegraphics[width=0.9\textwidth]{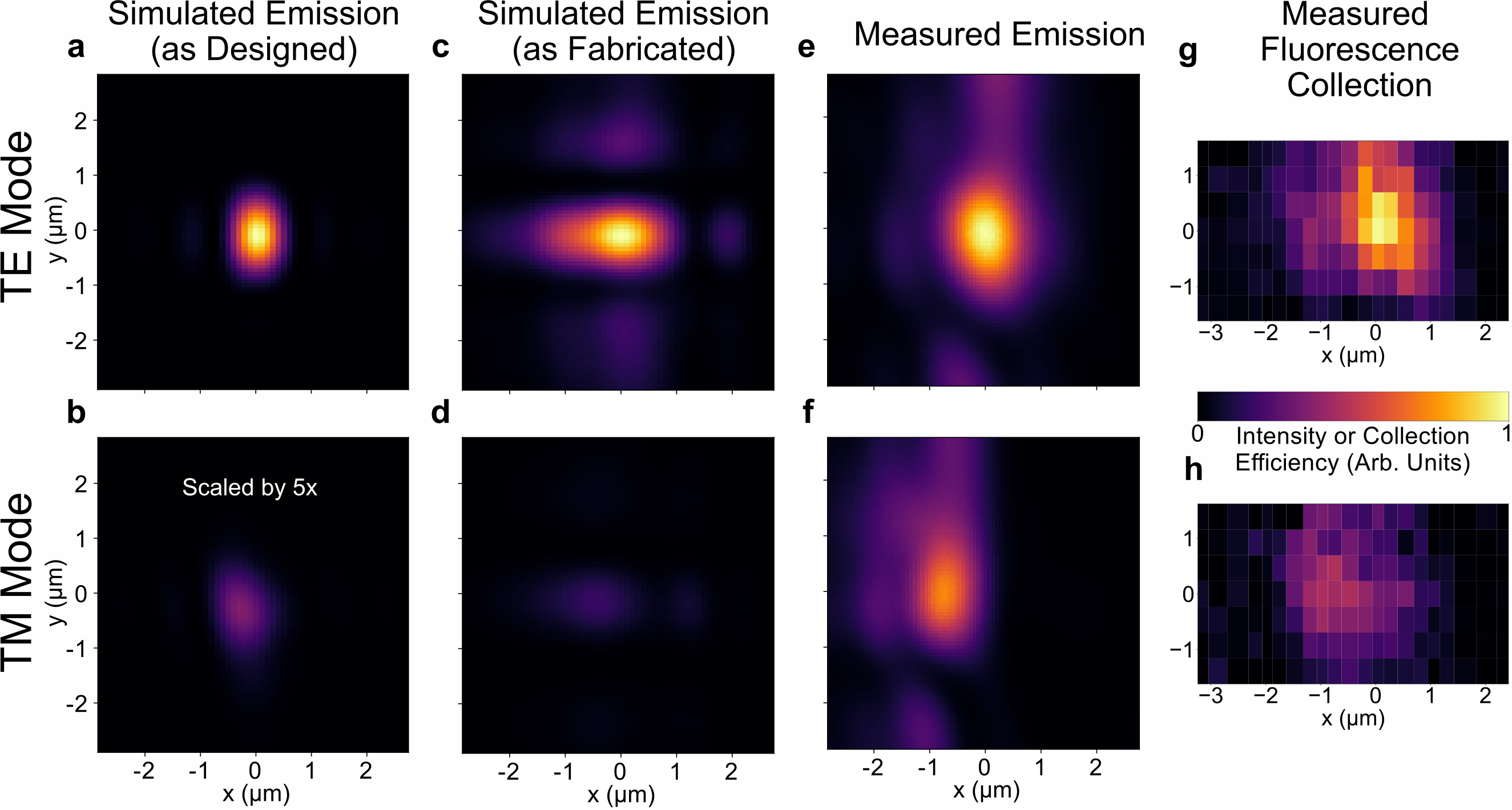}
\caption{Spatial profiles of grating emission and collection.  The columns depict the transverse-electric (TE) polarization mode (top row) and the transverse-magnetic (TM) polarization mode (bottom row) intensity profiles at the designed ion height (50 \textmu m) above the trap for the simulated emission of the ideal grating (\textbf{a},\textbf{b}), the simulation of the grating taking into account known fabrication defects (\textbf{c},\textbf{d}), the measured emission of the fabricated device (\textbf{e},\textbf{f}), and the measured ion fluorescence collection of the fabricated device (\textbf{g},\textbf{h}).  The intensity scale in each column is normalized to the maximum of the TE emission/collection, which is stronger (as designed) for this device. Note that the ideal grating TM mode intensity (\textbf{b}) has been scaled up by 5\(\times\) to make it clearly visible since the peak intensity of the TM mode for the ideal grating is only 7\% of the TE mode peak.  In all cases, the TE mode is displaced relative to the TM mode, as expected.
}
\label{fig:mode}
\end{figure}

The reciprocity of emission and collection allow us to characterize the expected fluorescence collection performance by profiling either the emitted or the collected mode (see Methods subsection~\ref{theory} for details on the calculation of the expected performance~\cite{PME,cornell}).  First, we verify the grating design using a 3D finite-difference time domain (FDTD) simulation of its emission profile, from which we calculate a 0.7\% expected collection efficiency, which is a substantial fraction of the 2.18\% solid-angle limit  (Fig.~\ref{fig:mode}a,b). Biases and unintentional alignment offsets in fabrication caused significant deviation from the designed geometry. We investigate the known discrepancies between the fabricated result and the ideal design using a second round of 3D FDTD and find an additional -8 dB loss, resulting in an expected collection efficiency of 0.11\% (Fig.~\ref{fig:mode}c,d). See Methods subsection~\ref{fab_modeling} for a detailed analysis. Next, we use an imaging system to profile the beam emitted by the fabricated device. Based on cross-sectional images of the beam at the target ion height, we calculate an expected collection efficiency of 0.041\(\pm0.007\)\% or -12.35 dB from the designed performance (Fig.~\ref{fig:mode}e,f).

Finally, we trap an ion over the collection grating and displace the ion to spatially map the fluorescence collection efficiency of the device (Fig.~\ref{fig:mode}g,h). A single-mode fiber coupled to the waveguide at the chip edge routes the collected light to a photodetector (photo-multiplier tube, PMT). To determine the integrated detection efficiency we measure the efficiency of the traditional, free-space detection pathway using a high NA compound lens outside the vacuum system and multiply it by the ratio of photon counts measured on the integrated and free-space detection pathways. Using the measured detection efficiency and accounting for detector and transmission losses, we determine a single-mode collection efficiency of 0.043\(\pm0.007\)\% for the integrated collection grating. This value agrees with the calculation based on the emission profile within experimental uncertainty. See the Methods subsections~\ref{characterization} and~\ref{loss} for a detailed accounting of the procedure and loss contributions.  

We additionally characterize the transverse-electric (TE) and transverse-magnetic (TM) polarization overlap and the resulting crosstalk to determine the collection grating's suitability for state detection and photon-mediated entanglement. For many atomic species, both polarizations are emitted equally in state detection; an optimally designed grating would thus maximize the strength of both the TE and TM emission and minimize their spatial separation since the optimal ion position lies at the maximum of the overlap. On the other hand, photon-mediated entanglement allows a magnetic field direction to be chosen that maximizes the desired ion emission into a definite polarization mode of the grating. For entanglement case, minimizing the overlap minimizes the incidence of error-inducing crosstalk. Our FDTD simulations show low crosstalk even though the maxima of the TE and TM modes are only 0.3 \textmu m (designed) and 0.5 \textmu m (as fabricated) apart since the grating emits much less of the TM mode (TM/TE ratio designed: 0.12; as fabricated: 0.18). We calculate a 13 dB suppression of TM crosstalk at the maximum of the TE emission for the designed grating and 9 dB for the simulation of the fabricated grating. The simulations thus show promise for photon-mediated entanglement, especially since additional photonic structures can effectively shed unwanted TM crosstalk~\cite{PME}. We find that the ratio of power diffracted for the TM mode versus the TE mode is larger in measurements than in simulation (measured TM/TE ratio: 0.69). Therefore, our measurements of the device show greater crosstalk for both ion collection and emission. Here, we measure the TM crosstalk at the optimal location for TE to be -5.3 dB for both methods, while the offset between the maxima in \(x\) is 0.7 \textmu m for emission and 0.8 \textmu m for collection. 

%%%%%%%%%%%%%%%%%%%%%%%%%%%%%%%%%%%%%%%%%%%%%%%%%%%%%%%%%%%%%%%%%

\subsection{State detection}

\begin{figure}[t]
  \centering
\includegraphics[width=0.75\textwidth]{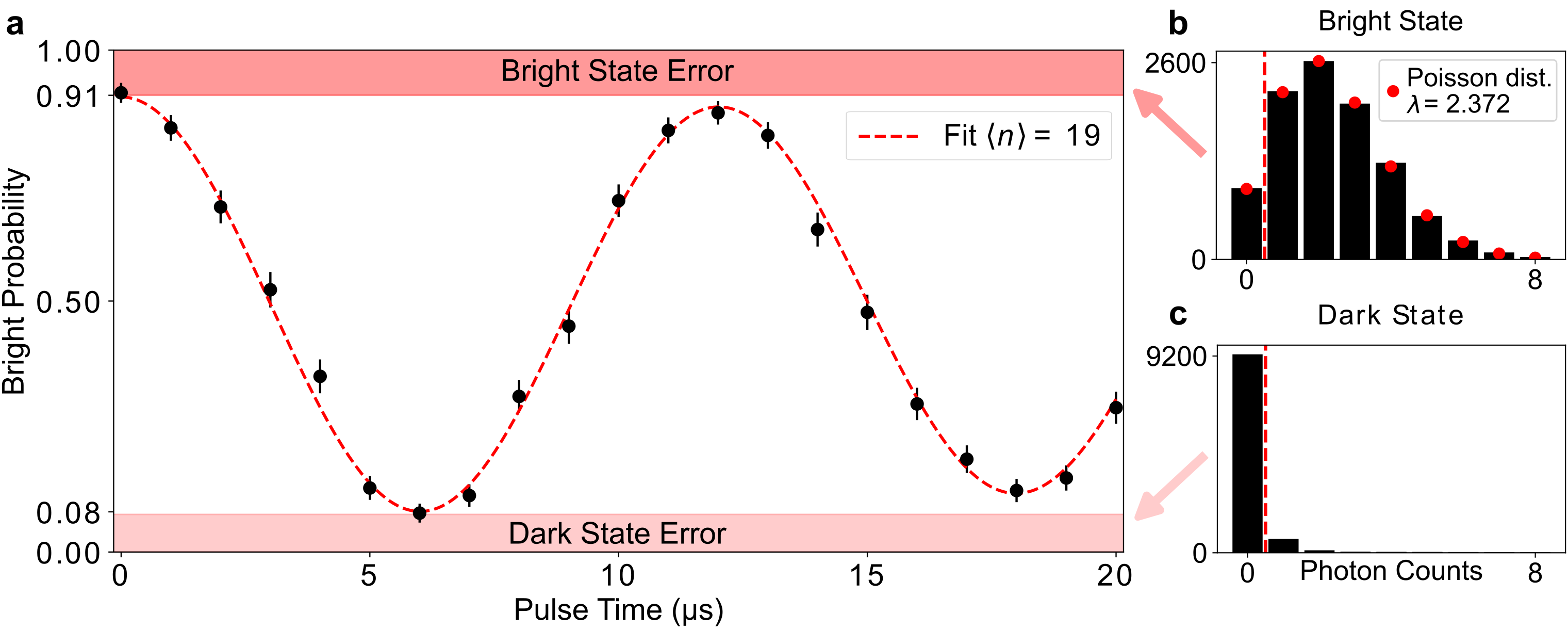}
\caption{Rabi oscillations of a Sr\(^{+}\) optical qubit measured via fluorescence collected using a tailored diffractive grating coupler. \textbf{a} We use the integrated collection pathway to determine the ion's state after excitation of the S\(_{1/2}\) to D\(_{5/2}\) transition at 674~nm. We plot the probability of measuring the optical qubit to be in the ground (bright) state as the length of the Rabi excitation pulse is scanned.  The excited state is not resonant with the readout light, so an ion in this state remains dark. Thermal motion of the ion causes the decay of the Rabi oscillations. The dashed line shows a fit assuming an average excitation of the lowest-frequency motional mode \(\langle n \rangle\) of 19~quanta, close to the Doppler-cooling limit.  Panels \textbf{b} and \textbf{c} show state detection histograms for bright (91\% fidelity) and dark (92.5\% fidelity) states. We classify a result as bright if we detect at least one photon in the 8 ms detection time.  The bright-state histogram is well-characterized by a Poisson distribution with mean \(\lambda=2.372\).}
\label{fig:state}
\end{figure}

As a proof of principle demonstration, we detect the state of a trapped ion by coupling the integrated collection device, via optical fiber, to an external photomultiplier tube. We distinguish between the \(\ket{1}=\ket{\text{S}_{1/2},m=-1/2}\) (bright) and \(\ket{0}=\ket{\text{D}_{5/2},m=-5/2}\) (dark) states by illuminating the ion with 422-nm light resonant with the \(\text{S}_{1/2}\rightarrow\text{P}_{1/2}\) transition in Sr\(^+\) that scatters rapidly in the bright state and does not scatter in the metastable dark state (Fig.~\ref{fig:overview}b). Figures~\ref{fig:state}b-c show the detection histogram for an ion initialized in the \(\text{S}_{1/2}\) level and for an ion shelved in the \(\ket{0}\) state, respectively, for an 8~ms detection time. The bright state follows the expected Poisson distribution, while the dark state distribution is non-Poissonian because it includes trials where the ion spontaneously decayed back to the \(\ket{1}\) state during the detection window (2\%) and trials where shelving failed (0.5\%). We use an 8 ms detection time with a one photon threshold to balance the bright signal against errors caused by dark counts and decay of the metastable state (\(\tau = 0.39\) s lifetime). These parameters allow us to detect the dark state with 92.5\(\pm\)0.3\% and the bright state with 90.7\(\pm\)0.3\% fidelity. We can detect the bright state faster using an adaptive protocol that classifies a trial as bright after the first count \cite{Oxford_MLE}. Splitting the 8 ms detection time into 10 bins allows us to classify a bright state in an average of 2.66 ms and thus an average of 5.33 ms to detect an equal mixture of bright and dark states. To complete our demonstration, we use the integrated pathway to measure the ion's state during excitation on the \(\ket{1}\rightarrow\ket{0}\) transition after cooling close to the Doppler limit and preparing in the \(\ket{1}\) state (Fig.~\ref{fig:state}a).

The achieved state detection duration and fidelity are limited by the observed detection efficiency. Of the -47.7 dB total photon loss, we attribute -16.6 dB to the solid angle subtended by the grating aperture, -17.1 dB to imperfect mode matching of the grating, -8.5 dB to transmission losses (dominated by \(\sim\)5.5 dB loss at the waveguide to fiber interface), and -5.5 dB to the detector's quantum efficiency (see Table~\ref{tab:loss} and Methods subsection~\ref{loss}). We observe a signal-to-background ratio of 36 with a bright rate of 297 s\(^{-1}\) and a dark rate of 8.1 s\(^{-1}\). The background count rate consists of 4.3 s\(^{-1}\) from scatter of the excitation laser, 3.0 s\(^{-1}\) detector dark counts, and 0.8 s\(^{-1}\) other background. Lowering transmission and detector losses would proportionally increase the signal and background from laser scatter, but improving the definition of the grating would increase both the signal and signal-to-background ratio since a smaller fraction of laser scatter would couple into the waveguide. 

% Define the shading color
\definecolor{lightgray}{gray}{0.9}

\begin{table}[t]
\centering
\begin{tabular}{@{} l c c c @{}}
 &
  \begin{tabular}[c]{@{}c@{}}Ion measurement \\ (dB)\end{tabular} &
  \begin{tabular}[c]{@{}c@{}}Emission-based \\ calculation (dB)\end{tabular} &
  \begin{tabular}[c]{@{}c@{}}With known \\ improvements (dB)\end{tabular} \\ \toprule
  \smallskip
\begin{tabular}[c]{@{}l@{}}Count ratio \\ (integrated to free-space)\end{tabular}  & -27.34\(\pm\)0.04  &    -         &      -      \\
\smallskip

\begin{tabular}[c]{@{}l@{}}Detection efficiency \\ (free-space)\end{tabular} & -20.34\(\pm\)0.1  &      -       &        -   \\
\smallskip

\begin{tabular}[c]{@{}l@{}}Loss from grating input \\ to detector\end{tabular}&   -    & -8.5\(\pm\)0.7    & -5       \\
\smallskip

\begin{tabular}[c]{@{}l@{}}Detector quantum \\ efficiency\end{tabular}        &    -   & -5.5 (PMT)  & -1.6 (SPAD) \\

\begin{tabular}[c]{@{}l@{}}Collection calculated \\ from emission profile\end{tabular} &    -   & -33.9 & -21.5             \\ \bottomrule

Total:     & -47.68\(\pm\)0.11 & -47.9\(\pm\)0.7     & -28.1       \\ 
\end{tabular}
\caption{Summary of photon losses. We measure the ion fluorescence collection by calibrating the efficiency of the traditional free-space detection and multiplying the result by the ratio of counts on the integrated and traditional collection pathways. The ion measurement agrees well with the expected performance based on the grating's emission profile and the routing and detection losses. See Methods subsection~\ref{loss} for a detailed explanation of the measurements and losses.  The right-most column displays the potential performance of the same device with improved fabrication fidelity, the use of a single-photon avalanche diode (SPAD) photon detector instead of a photo-multiplier tube (PMT), and the use of a multimode fiber instead of a single-mode PM fiber for coupling between the chip and detector.}
\label{tab:loss}
\end{table}

%%%%%%%%%%%%%%%%%%%%%%%%%%%%%%%%%%%%%%%%%%%%%%%%%%%%%%%%%%%%%%%%%

\section{Discussion}
In this work, we demonstrate the first collection of fluorescence from a trapped ion into a single-mode waveguide integrated into the trap chip. We introduce a grating design methodology that allows us to increase the effective solid-angle subtended by the collection grating and improve the grating's longitudinal focusing, and thus the collection performance. We fully characterize the designed and fabricated device's performance with numerical simulation, imaging of the grating emission, and profiling of ion fluorescence collection. This characterization allows us to experimentally validate the collection efficiency model presented in~\cite{cornell}. We model the effects of fabrication defects and experimental imperfections and understand their contributions to the detection efficiency (see Methods subsection~\ref{fab_modeling}). This understanding lays the foundation for dramatic improvement of the performance of future devices. We study the TE/TM mode overlap of both the design and the fabricated device, finding that the design suppresses crosstalk as desired for use in photon-mediated entanglement protocols~\cite{PME,oxford_reg}. Finally, we use the integrated collection pathway to measure the electronic state of an ion. The low background level in our measurement verifies the potential for this pathway to have the signal-to-noise ratio required for high-fidelity state detection. 

The verification of our design and the validation of our collection models allows us to project performance with straightforward improvements to address current limitations of the device. Fabricating the grating with the designed layer heights and features would increase the collection efficiency by up to 12 dB (see Methods subsection~\ref{fab_modeling} for details). Using a single-photon avalanche diode with 70\% quantum efficiency (QE) instead of a photo-multiplier tube (PMT) with 28\% QE would increase the detection rate by 4 dB. Using a multimode fiber to couple photons from the waveguide to a detector could eliminate 3.5 dB or more of the current \(\sim\)5.5 dB facet loss. These potential improvements to the current device could increase the photon detection efficiency by 19.5 dB without any design changes (see Table~\ref{tab:loss}). Scaling the fluorescence count rates by the full 19.5 dB and non-detector background count rates only by the 7.5 dB from routing and detection improvement would result in a detection time under 350 \textmu s with over 0.999 fidelity. For photon-mediated entanglement the relevant number is the projected total detection efficiency of -28.1 dB (see Table~\ref{tab:loss}).  For typical atomic-physics parameters, and excitation rates of approximately 1~MHz, this detection efficiency would lead to a few coincidences per second~\cite{Monroe2007,oxford_reg,PME}, without any further optimization of the grating or system design.    

Future designs could further improve performance through increased grating size and higher-resolution lithography to create a tighter focus and thus more efficient mode matching. Tiling multiple gratings in readout or entanglement-generation zones could further increase the solid angle subtended~\cite{PME}. The integration of waveguide-coupled detectors could eliminate facet-coupling losses and achieve QE approaching unity~\cite{waveguide_snspd}.  Furthermore, future designs may be tailored for either state determination or photon-mediated entanglement by exploiting the different requirements for polarization or laser-scatter to background discrimination. Together, these improvements have the potential to increase the probability to detect an emitted photon close to the solid angle limit of a grating subtending a large area below the ion and thus increase the expected photon-mediated entanglement rates beyond the level shown in recent free-space demonstrations. Photonic integration would additionally enable more straightforward incorporation of spatial multiplexing via arrays of ion pairs using photon-collection waveguide-optics.  Such multiplexing could allow for a near-linear increase in entanglement rate as a function of array size.  In summary, this proof-of-principle experiment demonstrates the feasibility of compact and modular fluorescence collection and coupling into integrated single-mode waveguides, with applications to state readout and photon-based entanglement in scalable trapped-ion quantum processors.

%%%%%%%%%%%%%%%%%%%%%%%%%%%%%

\section{Acknowledgments}
This material is based upon work supported by the U.S. Department of Energy, Office of Science, National Quantum Information Science Research Centers, Quantum Systems Accelerator. Additional support is acknowledged from the NSF Quantum Leap Challenge Institute for Hybrid Quantum Architectures and Networks (award no. 2016136) and through the Q-SEnSE Quantum Leap Challenge Institute (award no. 2016244). This material is based upon work supported by the Department of Energy and the Under Secretary of Defense for Research and Engineering under Air Force Contract no. FA8702-15-D-0001. Any opinions, findings, conclusions, or recommendations expressed in this material are those of the author(s) and do not necessarily reflect the views of the Department of Energy or the Under Secretary of Defense for Research and Engineering.

% MIT CQE (H98230-19-C-0292), NSF GRFP (1122374), MIT Locher Fellowships

\newpage

%%%%%%%%%%%%%%%%%%%%%%%%%%%%%%%%%%%%%%%%%%%%%%%%%%%%%%%%%%%%%%%%%

\section{Methods}

%%%%%%%%%%%%%%%%%%%%%%%%%%%%%%%%%%%%%%%%%%%%%%%%%%%%%%%%%%%%%%%%%

\subsection{Grating design}
Passive photonic systems are reciprocal. A diffraction grating designed to maximize the light emitted towards an ion will thus maximally collect light emitted by the ion~\cite{PME,cornell}. This allows us to interchangeably design and characterize the grating performance as either a focusing light emitter or, reciprocally, as a light collector. We design the focusing grating used in this demonstration by first identifying optimal grating geometries for fixed diffraction angles, using these optima to design a focusing solution in 2D, and then expanding this design in 3D. In this section, we provide a detailed description of an apodization technique, phase-shift apodization, that we introduce to vary the grating's diffraction strength and match the ion's emission intensity profile over the length of the grating.

%%%%%%%%%%%%%%%%%%%%%%%%%%%%%%%%%%%%%%%%%%%%%%%%%%%%%%%%%%%%%%%%%

\subsubsection{Diffraction grating parameterization and optimization}
\label{grating_parametrization}

To maximize the intensity of the emitted beam at the ion location, we design the grating to emit with a unidirectional, focused beam profile. Whereas a single-layer grating will radiate symmetrically, above and below the plane of the waveguide, the illustrated bilayer grating (Fig.~\ref{fig:grating_design}b) can be designed to preferentially radiate in one direction. In specific geometric cases, the diffraction in the downwards direction can be completely eliminated (e.g., offsetting the two layers of the grating by \(\frac{\lambda}{4}\) in both \(x\) and \(z\), where \(\lambda\) is the optical wavelength in the grating material). Complete suppression of diffraction in the undesired direction (down) is not possible for all combinations of wavelength and emission angle when the vertical bilayer separation is not a freely adjustable parameter, but directionality can be generally improved in any bilayer waveguide structure through optimal selection of the bilayer offset \(dx\) and the duty cycle in the upper (DCU) and lower (DCL) layers, as defined in Fig.~\ref{fig:grating_design}b~\cite{Notaros_bilayer2016}.

In addition to emitting unidirectionally, we desire the grating to focus at the ion location. To build a focusing grating, we first identify the range of angles needed to enable focusing in the propagation dimension (\(x\)) of the grating. Then, we construct a library of optimized geometric parameters (\(dx\), DCU, DCL) for upward emission at each of the angles within the required range. To find these optimal parameters, we iteratively simulate a short length of a fixed-period grating in 2D FDTD and use a particle-swarm algorithm to optimize the geometric parameters. The required 2D FDTD simulation volume is small, so typical simulation times are only a few seconds.

The particle-swarm algorithm maximizes the grating efficiency figure of merit \(\frac{\kappa}{\alpha + \kappa}\), with grating coupling coefficient into the desired upward mode \(\kappa\) and excess loss to undesired diffraction orders \(\alpha\). We derive \(\kappa\) and \(\alpha\) from \(P_T\), the total power lost from the grating waveguide over length \(L\), and \(P_D\), the power diffracted into the desired diffraction order. We implemented constraints that exclude duty cycle solutions resulting in feature sizes less than the minimum critical dimension attainable in our fabrication process (0.12 µm). Thus, as the pitch approaches 0.24 µm, this feature size constraint forces the duty cycle of both layers to 0.5, and only the bilayer \(x\) offset \(dx\) is truly optimized. However, generally the average duty cycle for the two layers is near 0.5.

%%%%%%%%%%%%%%%%%%%%%%%%%%%%%%%%%%%%%%%%%%%%%%%%%%%%%%%%%%%%%%%%%

\subsubsection{Modifying diffraction grating strength: phase-shift apodization}
\label{psa_descr}
The above simulations maximize the upward grating diffraction strength (\(\kappa\)) at a desired emission angle, and construct a suitably chirped grating that produces a focus, as illustrated in Fig.~\ref{fig:grating_design}c. However, maximizing the coupling efficiency also requires tailoring the intensity distribution of the diffracted light by varying the diffraction strength along the length of the grating.  In practice, this requires a mechanism known as apodization to locally reduce the coupling strength from its maximum value. Existing techniques for apodization include partial etching of the grating to various depths, adjustments to the duty cycle, and lowering the effective-index contrast via sub-wavelength patterning. The small features or varying etch depths used in these approaches typically require non-standard fabrication techniques~\cite{apodized_grating_design,apodized_grating_alt}. 

\begin{figure}[tb]
    \centering
    \includegraphics[width=0.7\linewidth]{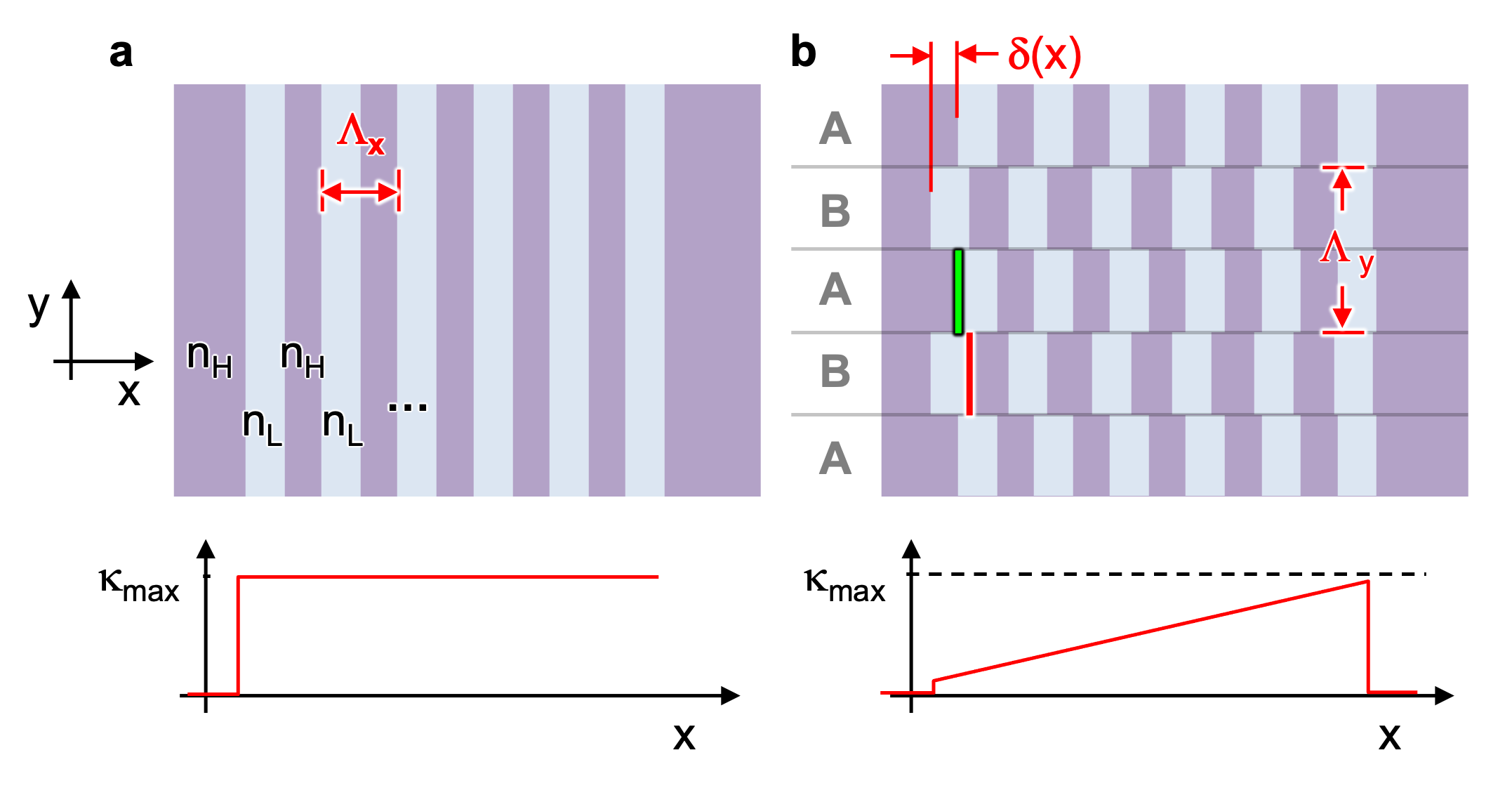}
    \caption{Schematic illustration of conventional, \textbf{a}, and phase-shift apodized, \textbf{b}, gratings. The grating coupler is comprised of materials with refractive indices \(n_H\) and \(n_L\) and period \(\Lambda_x\).  The lower panels show the grating strength \(\kappa\) as a function of position within the grating. The phase-shift apodized grating coupler has been subdivided into zones A and B with a sub-wavelength period \(\Lambda_y\). In a dielectric grating the leading edge (red) and trailing edge (green) of a grating line can produce scattering of equal magnitude but opposite phase. By shifting the grating features in zone B to be out of phase with those of A, net diffraction is reduced to zero since the combined scattering from two zones cancels out. Varying the phase shift thus continuously modifies the diffraction strength between zero and the maximum afforded by the materials' index contrast.}
    \label{fig:psa}
\end{figure}

Here we describe a new apodization technique based on phase shifts (Fig.~\ref{fig:psa}), where the grating is subdivided in the transverse (\(y\)) axis into two repeated zones (types A and B). This apodization technique is similar to that described in \cite{Simard2012,Greiner2008,apodized_grating_design} for narrow, single-mode waveguides. We adapt it to wide, large-area waveguides. The resulting shape of the apodized pattern is similar to the sub-wavelength patterns described in \cite{Luque-Gonzalez2021} used for dispersion engineering. Shifting the features in one zone along the propagation axis by an amount \(\delta\) changes the interference with light scattered from neighboring zones. This interference effect results in a tunable reduction of the diffraction strength by shifting between completely constructive and destructive interference. The only fundamental requirement is that the period of the zones in the \(y\) dimension \(\Lambda_y\) must be sub-wavelength (\(\Lambda_y < \lambda_m\)) to prevent unintended diffraction in the \(yz\) plane. Note that the grating's effective index shortens the optical wavelength in the grating waveguide \(\lambda_m = \frac{\lambda_0}{n_m}\). Using this apodization technique, we can continuously vary the diffraction strength from zero to the maximum afforded by the refractive index contrast of the waveguide materials with feature sizes of just under \(\lambda_m/2\) in a single binary etch pattern. Although the phase-shift grating structure is intrinsically 3D, we model it in 2D using an effective index structure and repeat the optimization described in Section~\ref{grating_parametrization} for several discrete values of phase shift to create a library of optimal grating parameters for a 2D focusing grating.

%%%%%%%%%%%%%%%%%%%%%%%%%%%%%%%%%%%%%%%%%%%%%%%%%%%%%%%%%%%%%%%%%

\subsubsection{Longitudinal grating design}

For optimal coupling, we tailor the diffraction strength along the length of the grating \(\kappa(x)\) to produce a diffracted intensity profile that replicates the intensity profile \(I_{\text{ion}}(x)\) that would be produced in the inverse case of the ion radiating light on the grating surface. We can approximate the intensity profile diffracted from the grating waveguide along the propagation axis \(I_{\text{diffraction}}(x)\) as a continuous function of \(\kappa(x)\) and \(\alpha(x)\): 

\begin{equation}
I_{\text{diffraction}} = \kappa(x) \cdot e^{-[\kappa(x) + \alpha(x)]x}. \label{eqn:4a}
\end{equation}

\noindent A priori, the function \(\kappa(x)\) necessary to replicate \(I_{\text{ion}}\) is unknown. One can deduce that \(\kappa\) should be small at the leading edge of the grating (to ensure that some fraction of the light propagates to the end of the grating), and very large at the end of the grating (to minimize light leaving the end of the grating without diffracting). We can assign a smoothly varying ansatz for \(\kappa(x)\): 

\begin{equation}
\kappa(x) = ax^3 + bx^2 + cx + d + A e^{Bx}. \label{eqn:4b}
\end{equation}

\noindent  

\begin{figure}[t]
    \centering
    \includegraphics[width=1.0\linewidth]{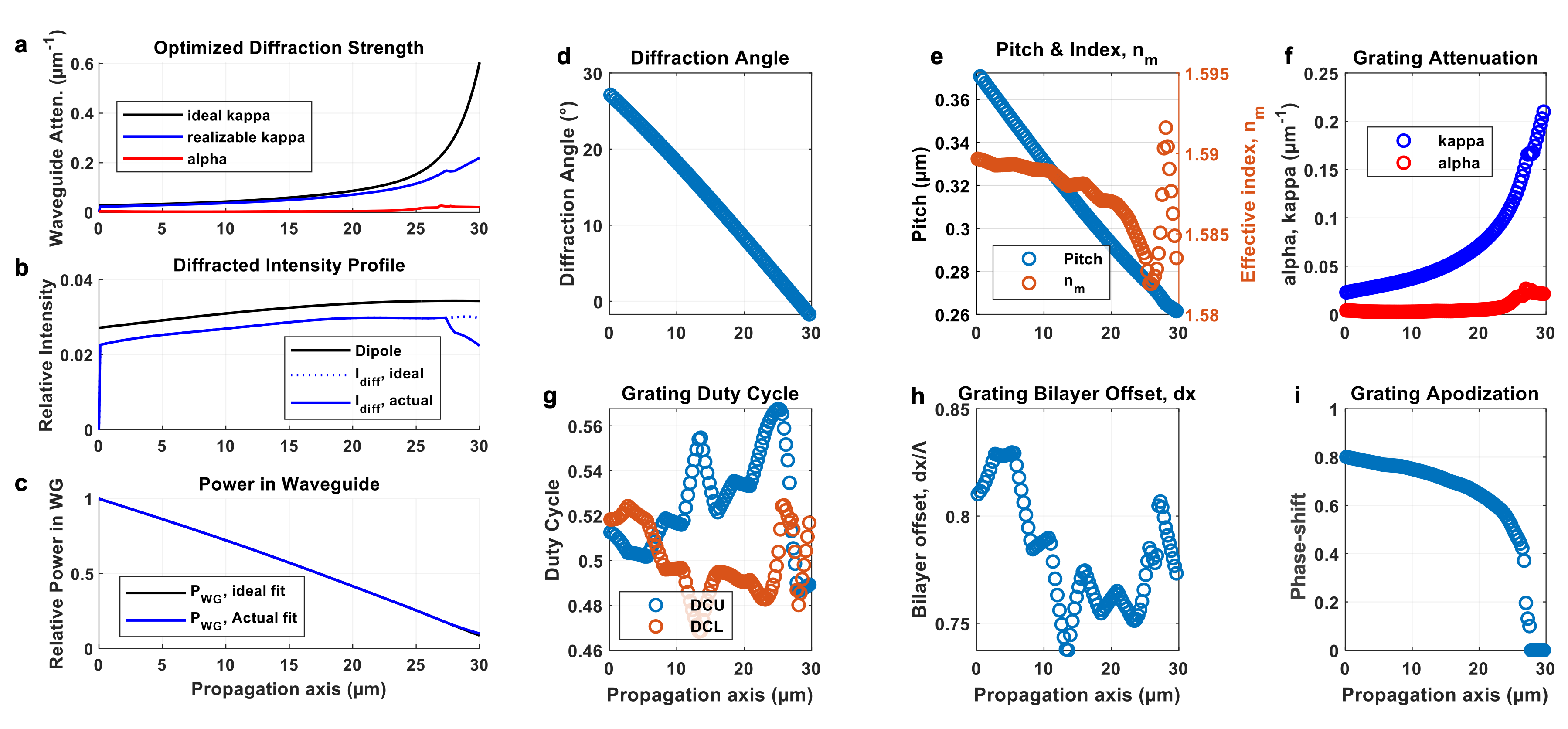}
    \caption{Optimized 2D-grating parameters for coupling to an ion trapped 50 \textmu m above the surface of a waveguide. \textbf{a} The library-constrained (red \& blue) grating loss coefficients closely match the ideal (black) for most of the grating and approximate the ion dipole intensity profile for \(x_{\text{ion}} = 28\) \textmu m (panel \textbf{b}). Panel \textbf{c} shows the success matching the ideal residual power propagating in the grating waveguide. The continuous solution of (\textbf{a}-\textbf{c}) is discretized to define a grating that varies from one cycle to the next to realize the desired diffraction angle \textbf{d}, requiring discrete pitches (\textbf{e}), and loss coefficients (\textbf{f}). The large ratio between \(\alpha\) and \(\kappa\) demonstrates the efficiency of the grating design. The optimized grating duty cycle and bilayer offset are given in (\textbf{g}-\textbf{h}) and contain jumps resulting from the particle swarm optimization used. The phase-shift to realize the variation in \(\kappa\) as shown in (\textbf{a}, \textbf{f}) is given in \textbf{i}.}
    \label{fig:2d_design}
\end{figure}

We use a least-squares difference between \(I_{\text{diffraction}}(x)\) and \(I_{\text{ion}}(x)\) to determine optimal values for the coefficients \(\{a, b, c, d, A, B\}\). For \(\alpha(x) = 0\), the optimal coefficients identify an ideal variation of \(\kappa\) along the length of the grating that maximizes coupling to the ion within the etendue limit afforded by the finite length of the grating. Then we identify the realizable optimal grating strength \(\kappa(x)\) by repeating the least-squares optimization subject to the constraints of non-zero excess loss \(\alpha\) and finite diffraction strength \(\kappa_{max}\). We obtain these parameters from the optimized grating parameter libraries described in the preceding section (Fig.~\ref{fig:2d_design}a). 

We combine longitudinal focusing and apodization using a lookup process. The process starts with the calculation of the required diffraction angle at a point \(x\) within the length of the grating from Snell’s law or Fermat’s principle of least time. Then we determine the optimal local coupling strength from our numerical fit of Eq.~\ref{eqn:4b}. Using the determined diffraction angle and coupling strength at a given point, we interpolate the grating pitch, optimum geometric parameters (DCU, DCL, \(dx\)), and phase-shift \(\delta\) from discrete solutions in the optimized grating parameter library. To define an actual grating we discretize the continuously varying solutions at each grating tooth (see Fig.~\ref{fig:2d_design}).

%%%%%%%%%%%%%%%%%%%%%%%%%%%%%%%%%%%%%%%%%%%%%%%%%%%%%%%%%%%%%%%%%

\begin{figure}[tb!]
    \centering
    \includegraphics[width=0.75\linewidth]{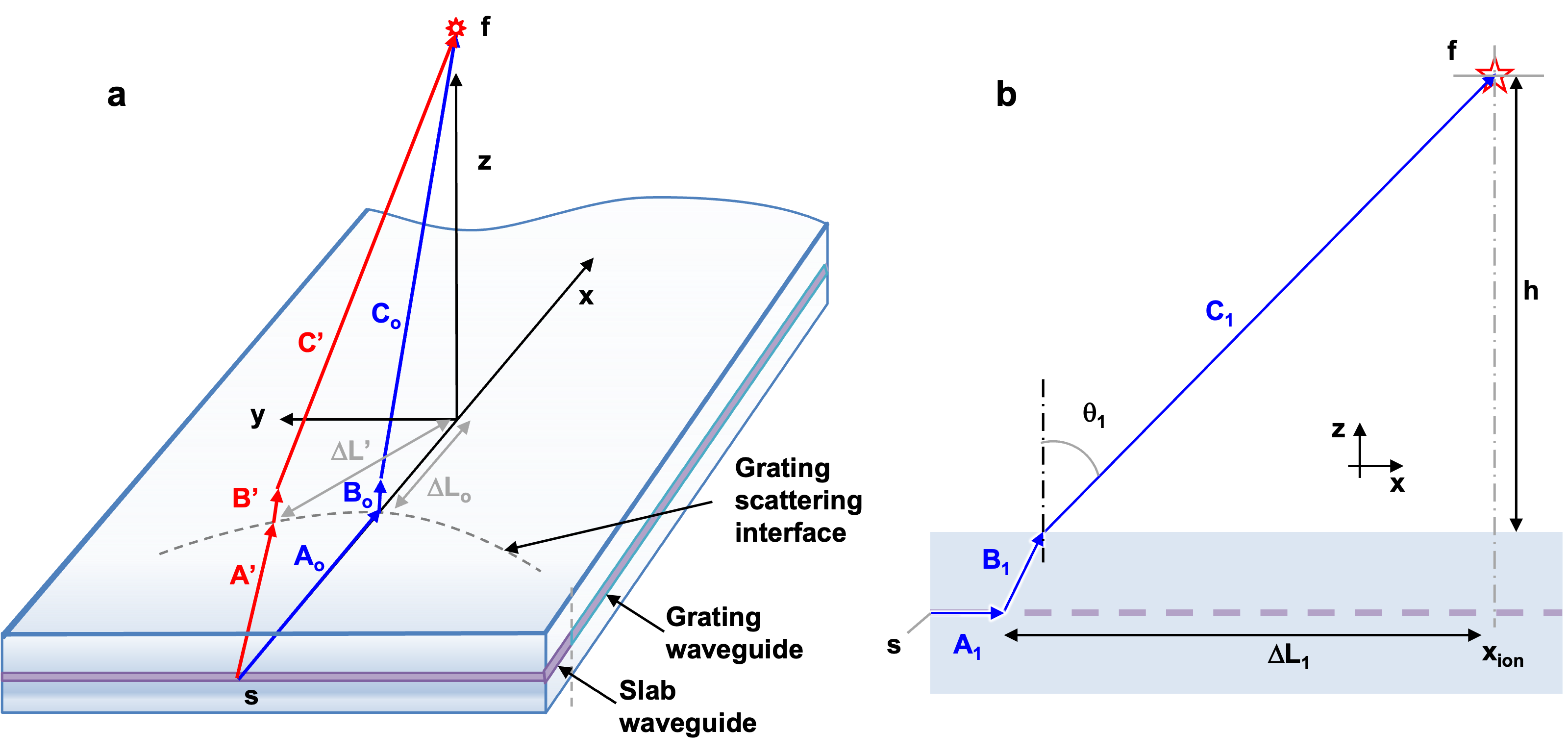}
    \caption{Geometry for defining the curvature of grating lines for a 3D focusing grating waveguide. \textbf{a} Perspective view and \textbf{b} cross-sectional view. The curvature of each grating scattering interface in the \(xy\) plane is defined by setting the optical path length to be the same in both paths \(A_oB_oC_o\) (blue) and \(A'B'C'\) (red), so that a photon transiting from ‘s’ to ‘f’ on any path arrives at the same phase.}
    \label{fig:3d_design}
\end{figure}

\subsubsection{3D focusing}
We realize focusing transverse to the axis of propagation by curving the grating lines in the \(xy\)-plane. We define the curved grating pattern using two steps. First, the previously defined 2D grating solution for longitudinal focusing and apodization sets the location of grating interfaces on the \(x\) axis. Second, the scattering interfaces are extruded in the \(y\) direction along curves that ensure photons from a common source remain in phase at the focal point (see Fig.~\ref{fig:3d_design}). Photons transiting the marginal path A'B'C' (red) travel a longer physical length, but segment A' in the slab waveguide (where the speed of propagation is slower) is shorter, such that photons transiting either ABC path arrive at ‘f’ in-phase. We derive the length of segments B and C from Snell’s law and the in-plane distance \(\Delta L\). The cumulative phase in transiting path BC is \(2\pi \cdot (n_{\text{SiO}_2} \cdot B + n_{\text{air}} \cdot C) / \lambda_0\), with \(n_i\) the refractive index of each material~\(i\). 

For the special case where the source at 's' has a flat phase front with a \(y\) width \(\gg \lambda_m\) (an expanded collimated beam), the cumulative phase in path A is approximately invariant in y, and is simply \(2\pi \cdot n_{\text{slab}} \cdot (x - x_s) / \lambda_0\), where \(x_s\) is the \(x\) location of the source and \(n_{\text{slab}}\) is the effective modal propagation constant of the slab waveguide. In a more generic case, where the phase varies in \(y\) (a common case being light diverging in the plane from a narrow single-mode waveguide), the cumulative phase in path A can be determined by simulating the propagation of light in a uniform planar slab waveguide emanating from a source at ‘s’ in 3D FDTD, and then extracting and unwrapping the phase \cite{Zhao2018} of the electric field propagating in the plane. For most cases, we find that the refraction at the interface of the slab waveguide and the grating waveguide is sufficiently small such that it can be neglected in calculating the grating curvature, though this could be accounted for by recalculating the in-plane phase map in 3D FDTD after the curvature of the first grating line that defines the interface has been determined.

\begin{figure}[tb!]
    \centering
    \includegraphics[width=0.7\linewidth]{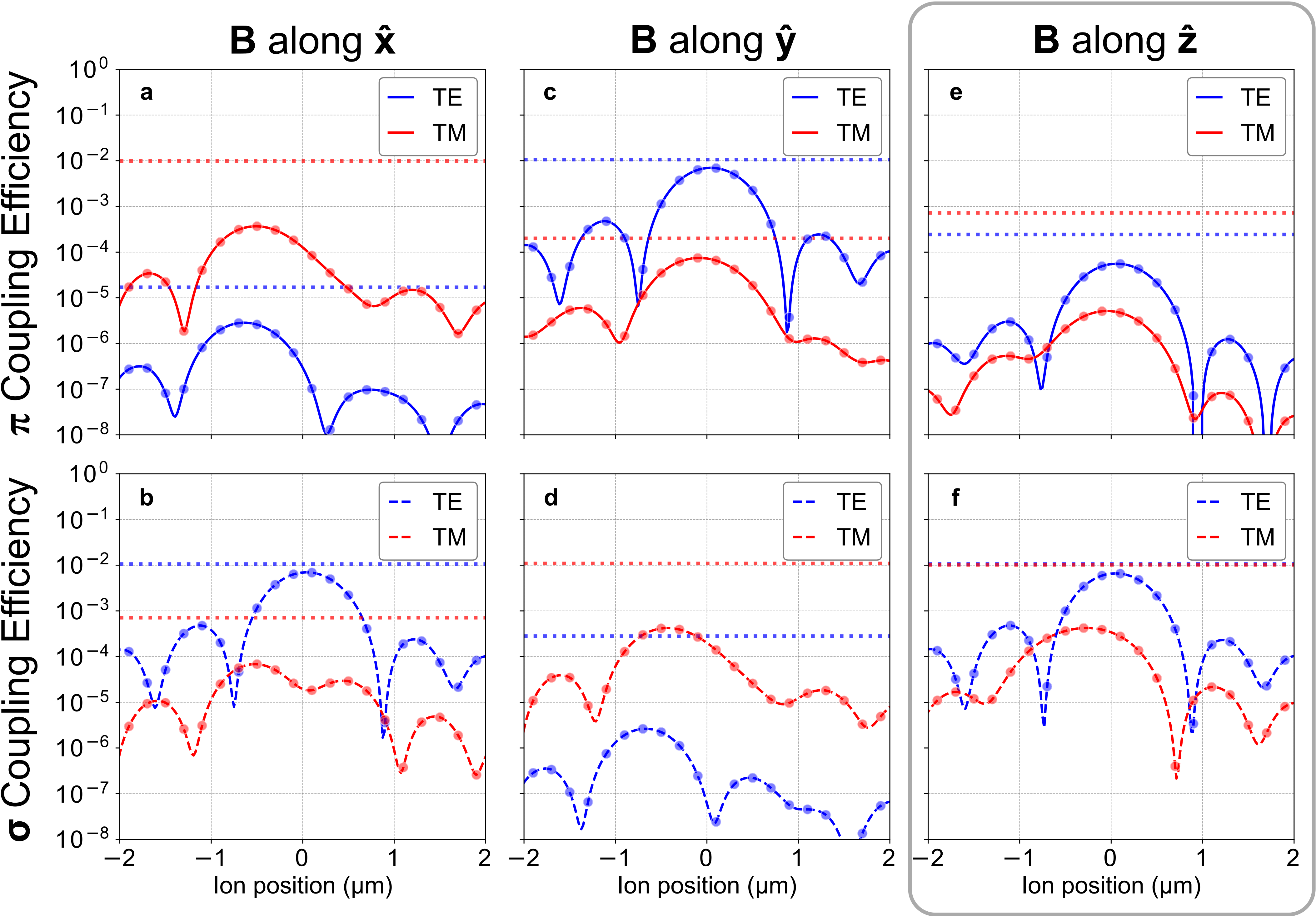}
    \caption{Simulations of ion emission coupling efficiency vs. ion displacement along the propagation axis. We investigate magnetic fields aligning the quantization axis along \(\hat{x}\) (\textbf{a}, \textbf{b}), \(\hat{y}\) (\textbf{c}, \textbf{d}), and \(\hat{z}\) (\textbf{e}, \textbf{f}). Points are simulated data and lines are interpolated splines to guide the eye. We highlight the \(\hat{z}\)-oriented magnetic field as that is the configuration used in our demonstration. The top row shows the collection of \(\pi\) emission as a fraction of the total ion fluorescence and the bottom row shows the \(\sigma\) fraction. The peak coupling efficiency of \(\sim\)0.67\% for a given polarization approaches the solid-angle collection limit of \(\sim\)1.09\% for the fraction of ion radiation in a given polarization that lands within the \(30\times30\) \textmu m footprint of the grating.}
    \label{fig:3d_lumerical_sims}
\end{figure}

%%%%%%%%%%%%%%%%%%%%%%%%%%%%%%%%%%%%%%%%%%%%%%%%%%%%%%%%%%%%%%%%%

\subsection{Grating characterization methods}
\label{characterization}
As mentioned in the Results section, we use three methods to characterize the collection device: 3D finite-difference time domain (FDTD) simulation, imaging of the emitted mode using a high-NA microscope objective, and imaging by scanning the position of a trapped ion while detecting fluorescence. This section will explain these methods in detail.

To determine the expected performance of the collection grating, we simulate multiple grating geometries starting with the geometry dictated by the design process detailed in the previous section. For 3D FDTD grating simulations, we launch a single mode of either transverse electric (TE, top row of Fig.~\ref{fig:mode}) or transverse magnetic (TM, bottom row of Fig.~\ref{fig:mode}) polarization into the start of the 30\(\times\)30 \textmu m\(^2\) grating. We then record the field emitted upwards by the grating just above the chip surface, since simulation of the entire air/vacuum volume up to the target ion height would require significant resources. We use far-field projection to generate the mode images at the nominal ion height (Fig~\ref{fig:grating_design}c and~\ref{fig:mode}). We verify our calculation of the expected collection based on the normalized field profile (see Section~\ref{theory}) by calculating the overlap of the simulated grating field with the ion radiation field at a location just above the chip surface and find good agreement between the two methods. We show the results of this overlap calculation for different choices of the quantization axis in Fig.~\ref{fig:3d_lumerical_sims}.

Beyond the ideal-case simulation, we simulate a variety of grating and chip-surface geometries that are motivated by SEM images of the grating to account for fabrication-induced discrepancies in grating performance. In Fig.~\ref{fig:mode} we compare the performance between our ideal-grating simulation (as designed) and the simulation including the all of the fabrication effects we investigated (as fabricated). For the depicted ``as fabricated" simulation, we accounted for the addition of a thin indium-tin oxide (ITO) film deposited on the surface of the ion-trap chip as a transparent conductor to shield the ion from exposed dielectric material, deviations in the thicknesses of the cladding and guiding layers from the design, and geometric deformations of the grating teeth. Further details and analyses are provided in the next section.

\begin{figure}[t]
  \centering
\includegraphics[width=0.6\textwidth]{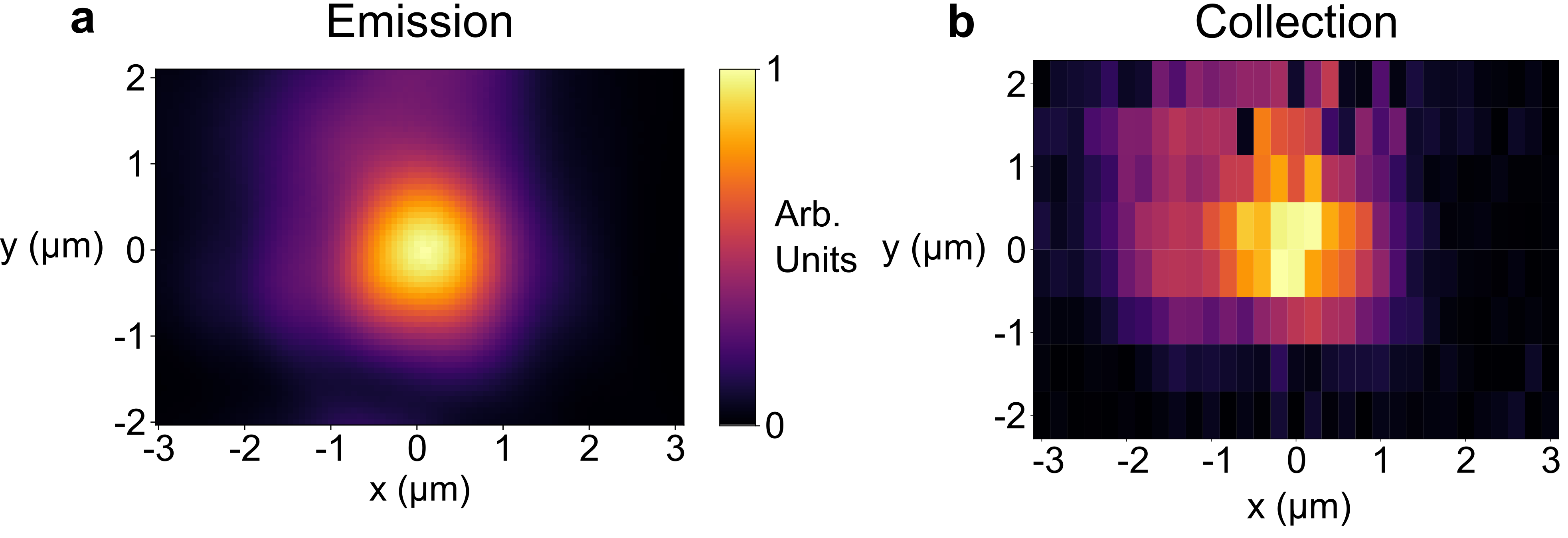}
\caption{The combined profile of TE and TM light emitted from the diffraction grating, \textbf{a}, matches the spatial profile of integrated fluorescence collection as we scan a trapped ion across the grating mode, \textbf{b}.
}
\label{fig:combined_modes}
\end{figure}

After fabrication, we measure the emission profile of the physical device. For this measurement, we send light into the grating from an edge-coupled polarization-maintaining (PM) fiber and image the emitted beam with a 50\(\times\), 0.95 NA microscope objective. By collecting images at a range of vertical offsets, we construct a 3D beam profile. To compare the TE and TM emission in relative intensity and position, we rotate the polarization coupled into the fiber by 90 degrees and re-measure the emitted beam without disturbing the imaging setup. Figures~\ref{fig:mode}e,f show a cross-section of the emitted beam at the nominal ion height for TE and TM light and Fig.~\ref{fig:combined_modes}a shows the scaled combination of the two. 

Once we have characterized the emission of the device, we measure its ion-fluorescence collection efficiency. We package the trap chip using a V-groove array of PM fibers that edge-couple to waveguides on the trap, and we install it in a cryogenic vacuum chamber as described in \cite{pgc_ions}. The PM fiber routes the collected light to a photo-multiplier tube (PMT) outside of the vacuum chamber for detection. We control the ion with light delivered by free-space optics. We measure the integrated fluorescence collection by driving the ion on the 422-nm transition and include a repumping laser beam at 1092 nm to make it a closed, cycling transition. We detect the emitted 422-nm photons using both the integrated pathway and our traditional imaging setup with a high-NA objective external to the vacuum system. By varying the DC electric potential, we raster the position of the ion in a plane at \(z=50\)~\textmu m to spatially map out the collection in \(x\) and \(y\). We plot the ratio of the count rate from the integrated pathway and the traditional pathway as a function of ion position in Fig.~\ref{fig:combined_modes}b. The RF axis of the ion trap is oriented along \(x\) and the displacement in \(y\) is limited by the increase of micromotion when the ion is displaced from the null. The dropouts at the top of Fig.~\ref{fig:combined_modes}b are caused by the ion's motion becoming unstable at these locations. To distinguish the TE and TM components of the collected light, we place a polarizer before the PMT and repeat the collection map while accepting polarizations corresponding to either TE or TM (Fig.~\ref{fig:mode}g,h).

\subsection{Modeling fabrication effects in simulation}
\label{fab_modeling}
As discussed in the Results section, the collection efficiency of the fabricated grating is approximately 12 dB lower than the designed efficiency. We therefore anticipate that we can substantially improve the efficiency by more faithfully reproducing the design in fabricated devices. To this end, we aim to quantify the impact of known deviations between our designed and fabricated chip geometries and, in turn, quantify the improvements we can expect from correcting for these deviations. We begin by simulating the ideal grating and chip geometry and sequentially introduce known deviations, re-simulating the cumulative effect each time to extract the effect of each deviation. We use the collection efficiency into the TE mode of the grating as our figure of merit, as it is the dominant collection contribution and obviates the need to run simulations for both TE and TM modes. For the designed geometry, we calculate a 0.67\% (-21.75 dB) TE collection efficiency (Fig.~\ref{fig:3d_lumerical_sims}f). 

First, we consider the surface of our trap chips. A 50-nm film of indium-tin oxide (ITO, a transparent conductor) covers the chip surface to shield the ion from the exposed SiO\(_2\), with a gap through the center to prevent shorting of the trap chip's electrodes. We shift the ITO in the \(y\) dimension of the grating by 1 \textmu m, to account for a measured offset between the chip's surface-metal and waveguiding layers. For this geometry, our simulated collection efficiency is -22.05 dB (\(\Delta_{\textrm{ITO}}=-0.3\) dB). This loss could be mitigated by reducing the ITO thickness. Next, we add a 500-nm-deep divot in the oxide within the ITO gap to account for a known divot introduced during fabrication of the chip's surface electrodes, resulting in a -24.95 dB simulated collection efficiency (\(\Delta_{\textrm{divot}}=-2.9\) dB). This divot was introduced during reworking of the electrodes during the fabrication process and can be avoided in the future.

In addition to these surface effects, we incorporate deformations of the grating layers into our simulations. Based on SEM images of several chips both during and after fabrication, we observe that the upper SiN layer thickness is typically reduced by 15 nm from design, the lower SiN layer thickness is typically reduced by 20 nm from design, and the SiO\(_2\) gap between the two layers is typically increased by 35 nm from design. Furthermore, we observe deformations of the phase-shift apodized grating-teeth that we assign to two categories: triangular and ellipsoidal (see Fig.~\ref{fig:overview}c). In the triangular-deformation case, the SiO\(_2\)-filled gaps defining the grating's teeth are under-etched during fabrication and turn out smaller and more triangular than designed. In the ellipsoidal-deformation case, these gaps are over-etched during fabrication and turn out larger and more elliptical than designed. We simulate both the triangular and ellipsoidal cases while also adjusting the layer and gap heights as discussed above. For the triangular case, the simulated collection efficiency is -27.25 dB (\(\Delta_{\textrm{tri}}=-2.3\) dB). We use the ellipsoidal case, with a simulated efficiency of -29.65 dB (\(\Delta_{\textrm{el}}=-4.7\) dB), for the grating in Fig.~\ref{fig:mode}e,f. For this grating, 63\% (-2 dB) of the input light is diffracted upward. We use this value to estimate the efficiency of the fabricated grating as discussed in the Results and in Section~\ref{loss}.

In sum (\(\Delta_{\textrm{ITO}}+\Delta_{\textrm{divot}}+\Delta_{\textrm{el}}\)), we can account for 7.9 dB of the approximately 12 dB performance difference between the designed and fabricated grating by incorporating known fabrication effects into our simulations. As a result, we anticipate that mitigating these fabrication effects would improve the performance of future collection gratings. We expect that the 4 dB discrepancy arises from effects that we were not able to characterize and are thus not yet incorporated into our simulations. Potential effects include uneven surface topography of the ITO which could help explain the transverse de-focusing, non-uniformity in the tooth deformations, as well as potential air gaps in the grating layers.

%%%%%%%%%%%%%%%%%%%%%%%%%%%%%%%%%%%%%%%%%%%%%%%%%%%%%%%%%%%%%%%%%

\subsection{Theoretical description of ion fluorescence collection}
\label{theory}

An atomic ion with a single valence electron that is excited to the \(\text{P}_{1/2}\) state has a 1/3 probability of decaying without changing the projection of angular momentum (\(\pi\) emission) and a 2/3 probability of decaying with a \(\Delta m=\pm1\) change in the angular momentum projection (\(\sigma^\pm\) emission). The combination of these two emission components is spherically symmetric in the case of unbiased excitation to the upper state (zero atomic polarization). However, the distribution of \(\pi\)- and \(\sigma\)-photons is inhomogeneous and follows the dipole emission pattern \(\bf{E}\propto\text{sin}(\theta)\bf{\hat{\theta}}\) for \(\pi\), while \(\sigma\) emission is distributed as \(\bf{E}\propto(\text{cos}(\theta)\bf{\hat{\theta}}\pm i \hat{\phi})\), where \(\theta\) is the polar angle and \(\phi\) the azimuthal angle from the quantization axis set by an external magnetic field. At any angle, the emitted light can be decomposed into equal contributions of two orthogonal linear polarizations with no radial component.

When calculating the total collection of ion fluorescence into the grating modes, the solid angle subtended by the grating gives a firm upper bound. Along the \(x\) axis, the ion is located 28 \textmu m from the start of the \(30\times30\) \textmu m\(^{2}\) grating. A thickness of 50 \textmu m of vacuum and 5 \textmu m of \(\text{SiO}_2\) cladding separate the ion from the grating in the \(z\) axis. The grating thus covers 2.18\% of the solid angle. A diffraction grating coupled to a single-mode waveguide can collect into either the quasi-TE or the quasi-TM mode of the waveguide. A maximum of 1.09\% could couple into each mode. For photon-mediated entanglement applications the magnetic field direction should be chosen to separate \(\pi\) and \(\sigma\) emission by maximizing the coupling into either the TE or TM mode and minimizing crosstalk from the other emission pattern into that mode. This strategy preserves the entanglement of an emitted photon coupled with the atomic state. Our demonstration uses a magnetic field oriented in the \(z\) direction, which results in predominantly \(\sigma\) emission (95.6\%) incident on the grating with near equal contributions to the TE and TM polarization directions (see Fig.~\ref{fig:3d_lumerical_sims}e,f).

The above description calculates the geometric limits on collection by a planar optic. To calculate the expected coupling efficiency (\(\eta\)) due to the mode overlap we use the formalism developed in~\cite{cornell}:
\begin{equation}
    \label{eq:Cornell}
    \eta = \frac{1}{16}\omega_0^2 |\mathbf{p}\dotproduct \mathbf{E}^*_g(\mathbf{r}_0)|^2,
\end{equation}
where \(\omega_0\) is the photon frequency, \(\mathbf{p}\) is the dipole polarization vector and \(\mathbf{E}^*_g(\mathbf{r}_0)\) is the electric field input into the grating normalized to unit power and evaluated at the position of the ion. The coupling is maximized by maximizing the projection of the normalized field onto the radiating dipole \(|\mathbf{p}\dotproduct \mathbf{E}^*_g(\mathbf{r}_0)|\).

To calculate the expected coupling given a measured beam intensity profile we scale the polarization vector to emit unit power (\(p_0 = \sqrt{\frac{3}{4\pi^3} \frac{\lambda^4}{c^3 \mu_0}}\)) [\(\frac{A m s}{\sqrt{W}}\)]~\cite{cornell} and express the grating electric field amplitude normalized to unit power \(E_g\) [\(\frac{V}{m\sqrt{W}}\)] in terms of the intensity per unit area normalized to unit power \(I_g\) [\(\frac{1}{m^2}\)]: 
\begin{equation}
    E_g = \sqrt{2c\mu_0}\sqrt{I_g}.
\end{equation}
Here \(\mu_0\) is the vacuum permeability. In our case of predominant \(\sigma\) emission incident on the grating, we can approximate the dot product \(|\hat{\mathbf{p}}\dotproduct\hat{\mathbf{E}}_g|\) as 2/3 from the Clebsch-Gordan coefficient from \(\sigma\) emission times 1/2 for the coupling to either the quasi-TE or quasi-TM mode. If the TE and TM intensity profiles are normalized by their respective total intensities and combined, then the maximum collection efficiency of Eq.~\ref{eq:Cornell} simplifies to 
\begin{equation}
    \eta = \frac{1}{4\pi}\lambda^2 \frac{\text{I}_{max}}{s^2},
    \label{eq:simple_collect}
\end{equation}
defined by a maximum normalized intensity at the brightest pixel of I\(_{max}\) [unit-less] for a pixel of length \(s\). We use this formula to calculate the expected coupling based on the simulated and measured beam profiles. We cross-checked Eq.~\ref{eq:simple_collect} with the full 3D FDTD simulation of the overlap between the ion emission and the grating mode and found a \(<\)2\% discrepancy, which we attribute to the assumption that all of the incident light is \(\sigma\) polarized.

%%%%%%%%%%%%%%%%%%%%%%%%%%

\subsection{Photon loss characterization}
\label{loss}
To characterize the loss mechanisms in the integrated collection pathway, we first measure the overall detection efficiency by deterministically emitting single photons from the ion. Then we determine the different loss mechanisms which can be divided into the overlap of the photon with the grating mode, the losses incurred routing the photon from the grating to the detector, and the detector quantum efficiency of 28\% (or -5.5 dB) taken from the PMT manufacturer specification for 422-nm light. 

Due to the low efficiency, directly measuring the detection efficiency on the integrated pathway would be prohibitively slow, so we calibrate the traditional bulk-optics imaging system using deterministically emitted single photons from the ion and use that calibration to determine the integrated efficiency by the ratio of photon counts from the two pathways. To calibrate, we repeat the following procedure. Resonant 422-nm light pumps the ion into the metastable \(\text{D}_{3/2}\) state. A pulse of 1092-nm light clears the \(\text{D}_{3/2}\) state via the short-lived \(\text{P}_{1/2}\) state such that the ion deterministically emits a single 422-nm photon. During the time window of the clearing pulse, we monitor the photo-detector for the emitted photon collected via an F/1.1 compound lens outside the vacuum chamber window. We determine the background counts by immediately repeating the 1092-nm pulse and detection step. Dividing the background-subtracted photon counts by the number of trials returns the detection efficiency of the bulk-optics imaging system. To determine the detection efficiency of the integrated path we illuminate the ion with both 422- and 1092-nm light and measure the counts detected using both the traditional and integrated paths. Multiplying the known detection efficiency of the traditional imaging system (\((9.24\pm0.22)\times10^{-3}\) or -20.34 dB) by the ratio of counts between the two systems (\((1.85\pm0.02)\times10^{-3}\) or -27.34 dB) returns the detection efficiency of the integrated path (\((1.71 \pm 0.04) \times 10^{-5}\) or -47.68 dB) (Table~\ref{tab:loss}). We also use this ratio method to measure the grating collection modes plotted in Fig.~\ref{fig:mode}g,h and Fig.~\ref{fig:combined_modes}b.

The overlap of an emitted photon with the grating mode depends both on the fraction of the solid angle subtended by the grating and the mode-matching of the ion emission to the grating mode. We measure the combination of the two factors by imaging the beam emitted by the grating at the target ion height for both TE and TM input (Section~\ref{characterization}). Since the ion emits equally into both polarizations, we combine the two normalized intensity profiles and use the method described in the previous section to determine the expected collection fraction of 0.064\% or -31.9 dB. Since we can only measure the total intensity emitted upwards by the grating for normalization and not the light at the grating input, we subtract 2 dB from this calculated value to account for light that starts in the grating mode but is not diffracted upward (see Section~\ref{fab_modeling}). This subtraction allows more accurate comparison to the simulations of grating collection which include the loss from incomplete upward diffraction (see Table~\ref{tab:loss}), but does not introduce any error since we add the 2 dB to the measured routing losses.

We determine the routing losses by back propagating light from the detector end and measuring the power in the grating emission. Starting at the grating end, the losses are comprised of incomplete upward diffraction (\(\sim\)2 dB, see above), light lost in the non-linear taper from the wide grating mode to the single-mode waveguide (\(<\)0.5 dB), light lost in the layer transitions from SiN to alumina waveguides (\(<\)0.6 dB), transmission and bend losses in the waveguides (\(\sim\)1 dB), reflections, scatter, and mode-matching loss at the waveguide-to-fiber edge coupler (\(\sim\)5.5 dB), and, finally, transmission and mating sleeve loss in the PM fiber (\(\sim\)1 dB). These approximate values are derived from separate measurements and simulations. We used optical adhesive as an index matching material to reduce the facet losses by 3 dB from 8.5 dB to the 5.5 dB we quote. The total loss we measure by back-propagating light is 10.5\(\pm\)0.7 dB. We determine this loss at the cryogenic operating temperature using our EMCCD camera to image the grating emission. We calibrate the EMCCD camera image by first measuring the emission at room temperature and comparing it with the result from power meter. The dominant uncertainty in the loss measurement is the background on the power meter measurement caused by scatter at the fiber-waveguide interface. 

There are three clear areas to greatly improve the efficiency of the collection and detection device without requiring a design change. Improving the fabrication to match the designed grating would reduce the mode-matching losses by about 12 dB (Fig.~\ref{fig:mode} and Table~\ref{tab:loss}). Using a multimode fiber instead of PM fiber would greatly reduce the mode-matching losses at the waveguide-fiber interface. In similar systems we have seen the loss at this interface go down to 2 dB without index matching material, which would be a \(\sim\)3.5 dB improvement. Improving the facet surface and using index matching material could further lower the interface loss below 2 dB by reducing scatter and reflections. Finally, replacing the PMT (28\% quantum efficiency) with a single-photon avalanche diode (70\% quantum efficiency) would reduce the losses by an additional 4 dB. We did not use a single-photon avalanche diode in our demonstration because the PMT had 10\(\times\) lower dark counts and thus offered higher fidelity for state detection.

\bibliographystyle{ieeetr}
\bibliography{bib}

\end{document}